\documentclass[aps,twocolumn,superscriptaddress,prl]{revtex4}%%preprint,,twocolumn
\usepackage[ngerman,english,french]{babel}
\usepackage{mathptmx}
\usepackage{lmodern} % math, rm, ss, tt
\usepackage{ifpdf}
\usepackage{graphicx}
\usepackage{times}
\usepackage{amsmath}
\usepackage{amssymb}
\usepackage{units}
\usepackage{stmaryrd} %% "// sign" = \sslash

\usepackage{ifpdf}
\usepackage{graphicx}
\usepackage{times}
\usepackage{amsmath}
\usepackage{amssymb}
\usepackage{units}
\usepackage[utf8]{inputenc}
\usepackage[T1]{fontenc}
\usepackage{stmaryrd} %% "// sign" = \sslash
\usepackage{lineno,hyperref}
\usepackage{ulem}
\usepackage{xcolor}

\usepackage[ngerman,english,french]{babel}
\usepackage{mathptmx}
\usepackage{lmodern} % math, rm, ss, tt
\usepackage{ifpdf}
\usepackage{graphicx}
\usepackage{times}
\usepackage{amsmath}
\usepackage{amssymb}
\usepackage{units}
\usepackage{stmaryrd} %% "// sign" = \sslash
\usepackage{xcolor}

\begin{document}

  \title{Manipulation of the magnetic state of a porphyrin-based molecule on gold:
From Kondo to quantum nanomagnet via the charge fluctuation regime}
   
\author{Yingzheng Gao}
\affiliation
% \affiliation[ESPCI]
{Laboratoire de Physique et d’Étude des Matériaux (LPEM), ESPCI Paris, Université PSL, CNRS UMR8213, Sorbonne Université, 75005 Paris, France}
\author{Sergio Vlaic}

\affiliation
% \affiliation[ESPCI]
{Laboratoire de Physique et d’Étude des Matériaux (LPEM), ESPCI Paris, Université PSL, CNRS UMR8213, Sorbonne Université, 75005 Paris, France}

\author{Tommaso Gorni}
\affiliation
% \affiliation[ESPCI]
{Laboratoire de Physique et d’Étude des Matériaux (LPEM), ESPCI Paris, Université PSL, CNRS UMR8213, Sorbonne Université, 75005 Paris, France}

\author{Luca de'~Medici}
\affiliation
% \affiliation[ESPCI]
{Laboratoire de Physique et d’Étude des Matériaux (LPEM), ESPCI Paris, Université PSL, CNRS UMR8213, Sorbonne Université, 75005 Paris, France}

\author{Sylvain Clair}
\affiliation
% \affiliation[AMU]
{Aix Marseille Univ, CNRS, IM2NP, Marseille, France}

\author{Dimitri Roditchev}
% \affiliation[ESPCI]
\affiliation
{Laboratoire de Physique et d’Étude des Matériaux (LPEM), ESPCI Paris, Université PSL, CNRS UMR8213, Sorbonne Université, 75005 Paris, France}
% \alsoaffiliation[INSP]
\affiliation
{Institut des Nanosciences de Paris, Sorbonne Université, CNRS UMR7588, 75005 Paris, France}

\author{Stéphane Pons}
\email{stephane.pons@espci.fr}
\affiliation
% \affiliation[ESPCI]
{Laboratoire de Physique et d’Étude des Matériaux (LPEM), ESPCI Paris, Université PSL, CNRS UMR8213, Sorbonne Université, 75005 Paris, France}
% \phone{+33 1 40 794 575}

%\usepackage[mathlines]{lineno}
%\linenumbers

\date{\today}

% \keywords{American Chemical Society, \LaTeX}
   
\begin{abstract}
\textbf{By moving individual Fe-Porphyrin-based molecules with the tip of Scanning Tunneling Microscope in the vicinity of a Br-atom containing elbow of the herringbone-reconstructed Au(111), we reversibly and continuously control their magnetic state. Several regimes are obtained experimentally and explored theoretically: from the integer spin limit, through intermediate magnetic states with renormalized magnetic anisotropy, until the Kondo-screened regime, corresponding to a progressive increase of charge fluctuations and mixed valency due to an increase in the interaction of the molecular Fe states with the substrate Fermi sea. Our results open a route for the realization, tuning and experimental studies of novel quantum magnetic states in molecule-surface hybrids.}

\end{abstract}

\maketitle
\section{Keywords}
spin-flip excitation, spin states, magnetic anisotropy, Kondo, charge fluctuation, mixed-valence states, scanning tunneling microscopy, transition metal complexes

\section{Introduction}
Addressing and manipulating the spin state of molecular species at interfaces is a challenge that could greatly benefit spintronics\,\cite{KhajetooriansAA2015}, nanoelectronics\,\cite{KumarKS2017,MolnarG2018} and quantum electronics\,\cite{HeinrichBW2013} in the near future. When the valence of a magnetic molecule deposited on a surface is integer, the description of spin-polarized molecular orbitals can be done in the framework of the atomic limit in terms of crystal field and spin-orbit coupling: the spin state is simply interpreted as a quantum magnet for which the Hund's rule determines the fundamental magnetic state. However, the atomic limit is no longer valid in the mixed-valence regime which is the most general behavior of an interacting magnetic impurity with an electron bath. 
\begin{figure}
  \centering
  \includegraphics[width = 8 cm]{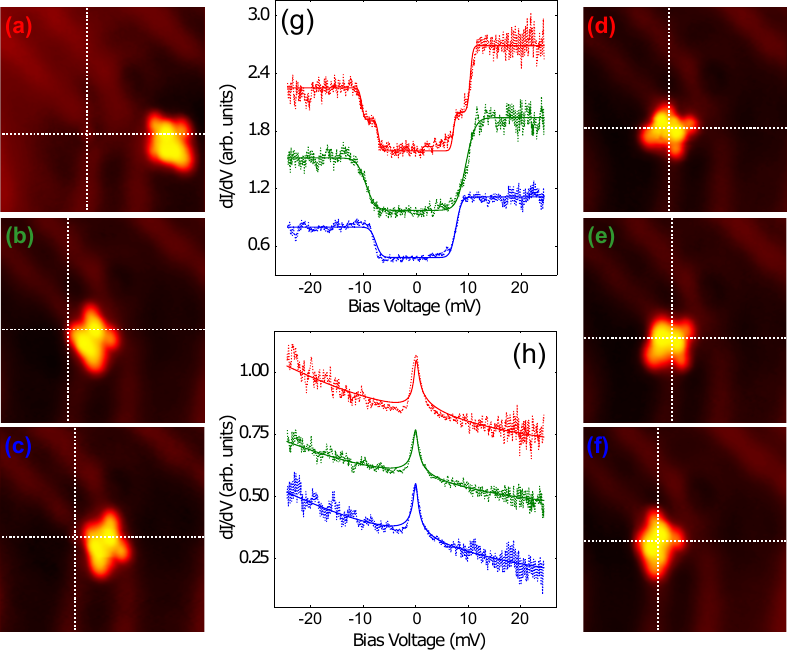}
  \caption{\textbf{Scanning tunneling study of $\rm{Fe-DPyDBrPP}$ monomers.}
\textbf{ (a-f)} Topography image of $\rm{Fe-DPyDBrPP}$ molecules at various location of the reconstructed surface of gold. The intersection of the dotted lines indicates the location of the Br adatom which decorates the elbow of the herringbone reconstruction. \textbf{(g)} Normalized scanning tunneling spectra taken on upper pyrroles in images (a-c) showing spin inelastic excitations depending on the location of the molecule on the surface.\,\textbf{(h)} Normalized scanning tunneling spectra taken on upper pyrroles in images (d-f), when the molecule is on top of the Br-decorated elbow site of the reconstruction. These spectra  show an Abrikosov-Suhl resonance witnessing a Kondo mechanism which does not depend on the in-plane rotation of the molecule. The solid lines correspond to the fits of the data (see Methods section). Experimental parameters: topography images: $V=125$\,mV, $I_{stab}=20$\,pA, size $10 \times 10$\,nm$^2$; spectroscopy: $V_{stab}=30$\,mV, $I_{stab}=200$\,pA, lock-in parameters: $V_{m}=0.2$\,mV, $f=750$\,Hz.}
  \label{fig:Singles}
\end{figure}

Among the most studied classes of magnetic molecules are the transition metal phthalocyanine (Pc) and porphyrin (P) families which are easily deposited at surface in vacuum and studied by various local probe microscopies and spectroscopies. In these molecules, the spin state is mainly given by the spin polarization of the central transition metal ion d-states\,\cite{tao2022structures,ZhangYY2011}. Several works have focused on the influence of external parameters on the magnetic ground state such as the influence of charge transfer to the orbitals of the molecule\,\cite{ ZhaoA2008,ChenX2008,KrullC2013,GarnicaM2014,ZhangQ2015,KaranS2016,OrmazaM2017,KaranS2018,GranetJ2020}, the effect of surface spin-orbit coupling and magnetic anisotropy\,\cite{HiraokaR2017,RubioVerduC2018}, the coupling to the substrate,\,\cite{HiraokaR2017,GirovskyJ2017, KuegelJ2018, BaljozovicM2021}, the interaction with attached and neighboring molecules\,\cite{IsvoranuC2011,WangY2017}, the structural deformation\,\cite{ParksJJ2010,LiR2018} or the chemical substitution of ligands\,\cite{HouJ2020}. 
For all these studies, a systematic understanding of the effect of mixed valence and charge fluctuations is still missing, although they have a strong influence on the effect of magnetic anisotropy.

Here we show that charge fluctuations in Fe 5,15-di-4-pyridyl-10,20-di-4-bromophenyl porphyrin ($\rm{Fe-DPyDBrPP}$) molecules adsorbed on the Br decorated Au(111) surface allow the magnetic state of the molecule to be driven between high-spin (S=1) and Kondo-screened states in a reversible and continuous manner through the intermediate valence regimes. 

\section{Results and discussion}
\subsection{Structural and spectroscopic properties of monomers}
$\rm{Fe-DPyDBrPP}$ were deposited in ultrahigh vacuum on Au(111) and studied in situ by scanning tunneling microscopy (STM) and spectroscopy (STS) at 1.3K (see Methods section for more details). The molecules are randomly located on hcp domains, fcc domains or stacking faults lines of the herringbone reconstruction\,\cite{BarthJV1990} of Au(111) and appear as bright spots in constant-current topographic STM images in figures\,\ref{fig:Singles} (a-f). The $\rm{Fe-DPyDBrPP}$ molecules do not exhibit a planar configuration on the Au(111) surface because two pyrrole rings bend toward the vacuum and exhibit a higher height in STM images; the other two pyrrole rings bend toward the substrate, resulting in a lower height in images. Such deformation corresponds to the saddling distortion of porphyrin-based molecules\,\cite{IshizukaT2022} that is quite commonly observed\,\cite{KimH2009,WangW2015,RubioVerduC2018,ChangMH2019,MengX2022}. Thus, the shape of the molecules in STM images allows us to determine the central position of the Fe-atom, the porphine macrocycle and the bromophenyl and pyridyl ligands attached to the macrocycle, as well as their location relative to the surface reconstruction. Importantly, during deposition and annealing, some Br-atoms detach from the molecules\,\cite{grill2007nano}, migrate at the surface, and get trapped by highly reactive elbows of the Au(111) herringbone reconstruction\,\cite{MeyerJA1996,ClairS2006}. The intersection of the dotted lines in figures\,\ref{fig:Singles} (a-c) points to the position of these elbows decorated by Br-adatoms.

Depending on their position on the surface, the molecules exhibit different spectral signatures. When a molecule is located on the hcp and fcc domains (figures\,\ref{fig:Singles} (a-c)), the tunneling spectra show conductance steps characteristic of a spin = 1 quantum magnet affected by the presence of magnetic anisotropy\,\cite{lambe1968molecular,HeinrichAJ2004,TsukaharaN2009,RubioVerduC2018,MengX2022}, figure\,\ref{fig:Singles} (g). The steps of conductance are due to the opening of additional spin-flip tunneling channels trough inelastic excitations. This is in agreement with the known behavior of Fe-Pc and Fe-P which behave as S=1 nanomagnets once deposited on the Au(111) surface\,\cite{Maki2012,HiraokaR2017,WangY2017,LiR2018,RubioVerduC2018,MengX2022}. The presence of single step or double step is a quite common phenomenon, e.g. ref.\,\cite{TsukaharaN2009,RubioVerduC2018,MengX2022} and supplementary information of ref.\,\cite{RubioVerduC2018}. The step-like spin-flip signatures are recorded at both upper pyrrole and Fe-atom locations, providing evidence for hybridization of the molecular states of the pyrrole with the Fe magnetic d-states\,\cite{RubioVerduC2018,MengX2022}. 

When the molecule is located on the elbow of the reconstruction, it behaves differently and exhibits a spectral resonance at the Fermi level. The molecule rotation induced by the microscope tip over the elbow site barely affects the shape and amplitude of the spectral resonance, figures\,\ref{fig:Singles} (d-f, h). 

\begin{figure}
  \centering
  \includegraphics[width = 7 cm]{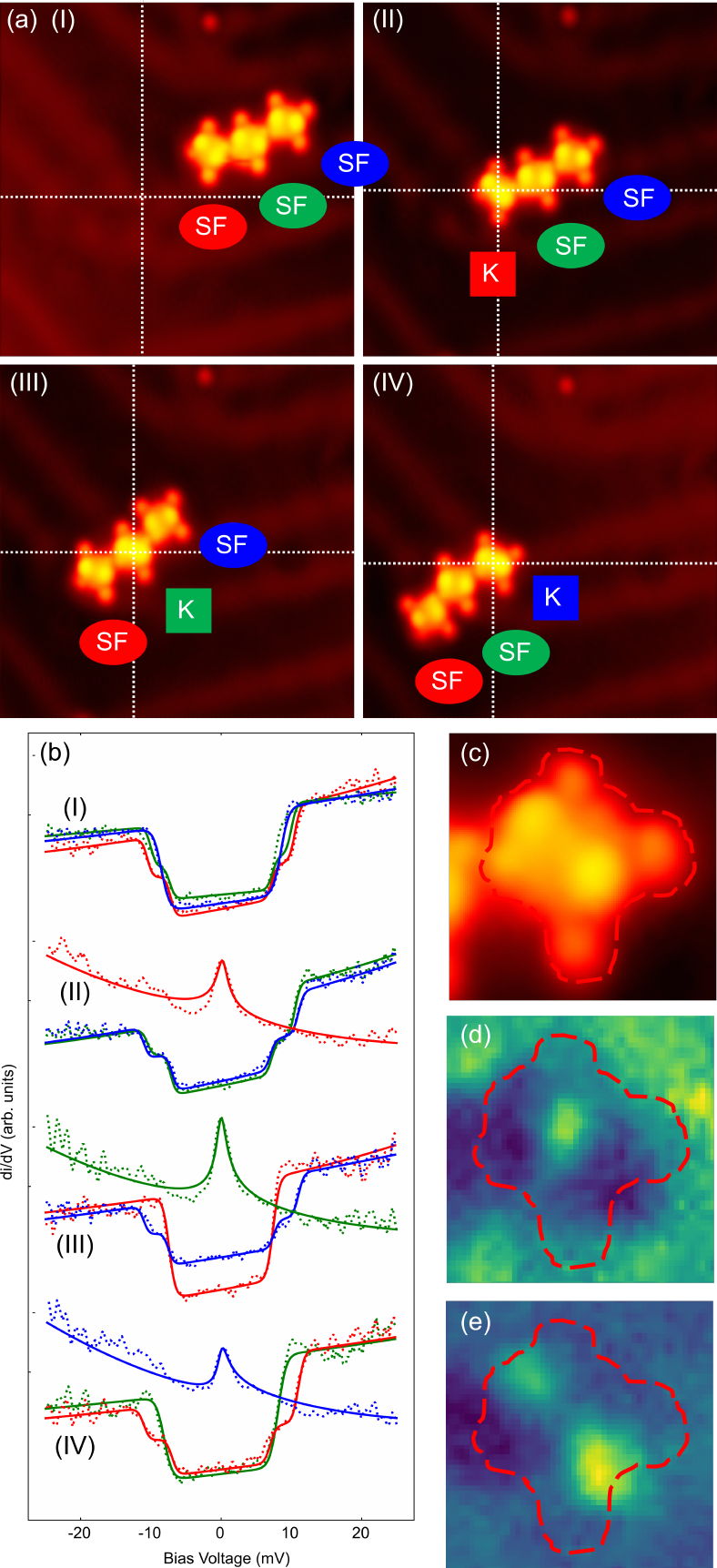}
  \caption{\textbf{Scanning tunneling study of $\rm{Fe-DPyDBrPP}$ chains made by Ullman's coupling.} \textbf{(a)} topography images (I-IV) of a trimer chain of molecules made by Ullman's coupling which is manipulated by the tip of the microscope in order to position sequentially the monomers above the Br-site. The intersection of the dotted lines indicates the Br-site. The molecules are labeled as a function of their spectroscopic signature measured in (b): "SF" stands for spin-flip and "K" for Kondo. Experimental parameters of images (I-IV): size $15 \times 15$\,nm$^{2}$, $V=125$\,mV, $I=20$\,pA.  \textbf{(b)} related scanning tunneling spectra taken above the left upper-pyrrole of each molecule of the chain, showing the presence of the Kondo peak for the molecules located above the Br-site which otherwise shows a spin-flip signature. V$_{stab}=30$\,mV, $I_{stab}=200$\,pA. Lock-in parameters: $V_{m}=0.2$\,mV, $f=750$\,Hz. The solid lines correspond to the fits of the data (see Methods section). \textbf{(c)} topographic zoom on a molecule of the chain exhibiting a spin-flip spectroscopic signature. \textbf{(d-e)} differential conductance image recorded simultaneously to image (c) at $-100$\,mV and $100$\,mV respectively showing the symmetry of the frontier orbitals below and above the Fermi level. The red dotted line delimits the molecule. Experimental parameters of (c-e), size $3\times3$\,nm$^2$, V$_{stab}=800$\,mV, $I_{stab}=500$\,pA. Lock-in parameters: $V_{m}=5$\,mV, $f=900$\,Hz.}
  \label{fig:Chains}
\end{figure}

In the following we show that the two distinct spectral signatures are fully controlled by the adsorption site of the molecule. To this end we have prepared chains of 3 covalently bonded $\rm{Fe-DPyDBrPP}$ molecules by Ulmann's coupling (see Methods section). In figure\,\ref{fig:Chains}, a trimer chain is moved by the tip of the microscope to various positions of the reconstructed surface. The targeted locations are the hcp and fcc domains and the Br-decorated elbow of the reconstruction. In the first panel (I) of figure\,\ref{fig:Chains}, the 3 molecules are located inside a fcc domain. In panels (II-IV), the molecular chain is sequentially repositioned with the microscope tip to move the iron center of each molecule over the Br-decorated site. When the molecules are located inside fcc and hcp domains, the spectra exhibit inelastic excitations of independent spin 1 nanomagnets in presence of magnetic anisotropy, similar to which was already measured for monomers in figure\,\ref{fig:Singles}. This observation means that, here, neither the nature of the covalent bonds nor the substrate mediated interaction are efficient enough for coupling the molecules together. The spectra usually display a symmetric double-step structure. Following the standard analysis\,\cite{HeinrichAJ2004}, the characteristic voltages of the steps is interpreted to be related to the out of plane and in plane magnetic anisotropy energies. Depending on adsorption sites, the out of plane and in plane magnetic anisotropy energies can vary from $6.8$ to $10.0$\,meV and from $0$ to $1.5$\,meV respectively. The out of plane anisotropy parameters correspond to typical values of Fe porphyrin and phthalocyanine based magnetic molecules adsorbed on gold\,\cite{TsukaharaN2009,LiR2018,HiraokaR2017,RubioVerduC2018,MengX2022}. The variation of the measured transverse anisotropy can be explained by the many possible configurations of the two fold symmetry $\rm{Fe-DPyDBrPP}$ molecules on the highly inhomogeneous Au(111) surface. 

In panels (II-IV), the molecules are moved one by one above a Br-decorated elbow. In each case, the molecule exhibits a spectral resonance at 0 bias similarly to the monomer in figure\,\ref{fig:Singles}. In figure\,\ref{fig:Chains}, the resonance is found to be reversible once the molecule is removed from the elbow site and independent of the selected molecule. We identify the spectral peak as an Abrikosov-Suhl resonance\,\cite{Abrikosov1965,Suhl1965,Nagaoka1965} (also named Kondo peak) due to the many body Kondo interactions of the Fe d states with the substrate electron bath. The characteristic Kondo temperature, $\rm{T_K} \approx 11$\,K, evaluated by fitting the lineshape with a Frota function\,\cite{frota1986,frota1992}, is found to be independent of the molecular orientation with respect to the surface atomic lattice. We therefore expect that magnetic anisotropy plays no role. 

The Kondo effect originating from degenerate triplet ground state is extremely sensitive to magnetic anisotropy which tends to lift the degeneracy\,\cite{ParksJJ2010}. In the present case, $\rm{K_BT_K} \approx 0.9$\,meV is much smaller than the measured anisotropy. Three possible phenomena can explain the robustness of the Kondo effect on the magnetic anisotropy occurring at the Br-decorated elbow site: 1. The charge fluctuations due to the coupling of the Fe states with the substrate states\,\cite{JacobD2018,ZitkoR2021,BlesioGG2023} overwhelm the magnetic anisotropy energy at this location, effectively restoring the degeneracy and screening the magnetic S=1 moment; 2. A sizable charge transfer from the substrate or the ligands onto the Fe atom induces a spin reduction from S = 1 to S = 1/2, resulting in a spin 1/2 Kondo effect which is naturally immune against magnetic anisotropy\,\cite{JacobD2018}; 3. The deformation of the molecule at this position induces the suppression of the magnetic anisotropy energy\,\cite{ParksJJ2010}. This last explanation is not in agreement with our tunneling topography studies which did not reveal a significant deformation of the molecule whose shape is preserved on the Br-site. It has also been shown that the anisotropy energy is not significantly affected when a similar molecule is pressed onto the surface with the tip of a microscope while staying away from the tip-molecule contact\,\cite{HiraokaR2017,KaranS2018,MengX2022}. Scenario 2, which assumes that charge transfer from the substrate gives rise to a molecular doublet, is the natural interpretation and has been analyzed in previous studies exploring different configurations\,\cite{OrmazaM2017,KaranS2018}. The experimental data, DFT calculations and theoretical interpretation that follow lean more towards the first scenario in the present case.

\subsection{Density Functional Theory analysis}
DFT was used to study the electronic properties as a function of the distance of the molecule from the substrate. The latter is not easily determined by standard state-of-the-art DFT calculations describing the electronic properties accurately because it involves a delicate equilibrium with Van der Waals forces\,\cite{HermannJ2017}. We have therefore used this distance as a parameter.
\begin{figure*}
\centering
\includegraphics[width=15 cm]{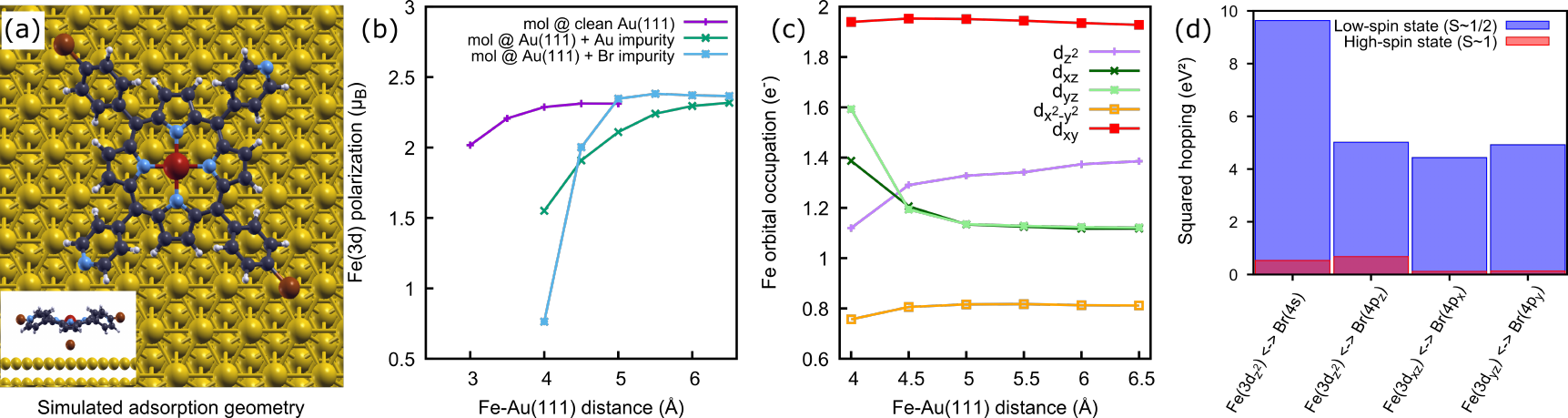}
\caption{\textbf{Theory analysis by DFT calculations.} \textbf{(a)} Simulated adsorption geometry for the Iron(II) 5,15-(di-4-bromophenyl)-10,20-(di-4-pyridyl) porphyrin molecule on a Au(111) surface. The (Au,Br) impurities are adsorbed on the hollow-hcp site. \textbf{(b)} polarizations  as a function of the distance from the substrate, computed for three different substrates, clean Au(111), Au(111) + a Au adatom and Au(111) + a Br adatom, obtained via the same Mulliken analysis as in (c). \textbf{(c)} Orbitally-resolved Fe(3d) charge as a function of the distance from the substrate covered with a Br adatom. The charge and magnetization values have been obtained via a Mulliken analysis projecting the Kohn-Sham orbitals over the atomic states of the Fe pseudopotential. \textbf{(d)} Calculated non-zero squared hoppings between the Fe(3d) and the Br(4s,4p) states of the Br adatom on Au(111) in the low-spin (Fe-Au(111) distance = $4.0$\,\AA, in blue) and high spin (Fe-Au(111) distance = $5.0$\,\AA, in red) case. Only the squared average of the spin-up and spin-down hopping is reported, since no significant difference has been found in the two channels. }
\label{fig:DFT1}
\end{figure*}

In order to rationalize the effect of the Br-adatom on the electronic properties of $\rm{Fe-DPyDBrPP}$, two sets of situations were simulated: 1. as a function of the distance to the genuine Au surface 2. as a function of the distance to a Br (or Au) adatom on Au slabs positioned below the Fe atom of the molecule $\rm{Fe-DPyDBrPP}$. 

The Fe 3d states are dominant in the coupling with the substrate through the Br states. Fe d$_{\rm xz}$, d$_{\rm yz}$ states are mainly hybridized with the pyrroles molecular orbitals. Therefore, further analysis will be discussed on the basis of the Fe 3d hybrids with molecular orbital. In particular the Fe d$_{\rm xz}$, d$_{\rm yz}$ and d$_{\rm z^2}$ hybrid states were proven to be at the origin of the observed phenomena. 

The DFT simulation shows that the magnetic polarization (contrary to the total charge) of the Fe atom, figure\,\ref{fig:DFT1} (b), depends strongly on the molecule-surface distance when the surface is decorated with a Br (or Au) adatom, whereas it is much less sensitive when approaching the clean surface. When the molecule is away from the Br site, the total occupancy and magnetic moment are about 6.4 $e^-$ and $2\,\mu_B$ which corresponds to a configuration close to the spin 1 state (high spin state) and the hybridization of the Fe d states with the substrate is small. This is in good agreement with the measured spectroscopic signature of a spin 1 quantum magnet when the molecule is above fcc or hcp domains. Indeed, the lack of a Kondo signature may be the result of weak hopping integrals from Fe orbitals to surface orbitals when the molecule is far away. Hybridization with the substrate cannot compete with the magnetic anisotropy and the Kondo effect is prevented. The magnetic anisotropy energy dominates the physics and the experimental spectroscopic signature is that of a spin 1 subjected to an anisotropy of a few meV. 

Moving the molecule of 1\,\AA\,towards the Br adatom, from 5 to 4\,\AA, leaves the occupancy of the Fe states roughly unchanged (only a minor reshuffling of the charge distribution between the orbitals is observed (figure\,\ref{fig:DFT1} (c)) - but induces a strong reduction of the magnetic polarization from about $2\,\mu_B$ to about $1\,\mu_B$. The approximately constant charge suggests that charge fluctuations are the cause of this reduction, which is confirmed by the fact that the hybridization of the Fe and Br states increases exponentially as the molecule is moved closer to the surface (the squared hoppings that enters in this form in the hybridization function between the impurity and the substrate are plotted in the figure\,\ref{fig:DFT1} (d)). The squared hopping integrals of the Fe states to the surface are indeed one order of magnitude smaller when the molecule is far away. The Fe 3d$_{z^2}$ hybrids and Br 4s states, which have a almost perfect rotational symmetry, are supposed to dominate the physics when the molecule is closer (figure\,\ref{fig:DFT1} (d)). This configuration is favorable to the emergence of a charge transport that is insensitive to magnetic anisotropy and independent of the relative orientation between the molecule and the surface once the iron atom is located just above the adsorbed Br atom, in perfect agreement with observations.

\begin{figure*}
  \centering
  \includegraphics[width = 15 cm]{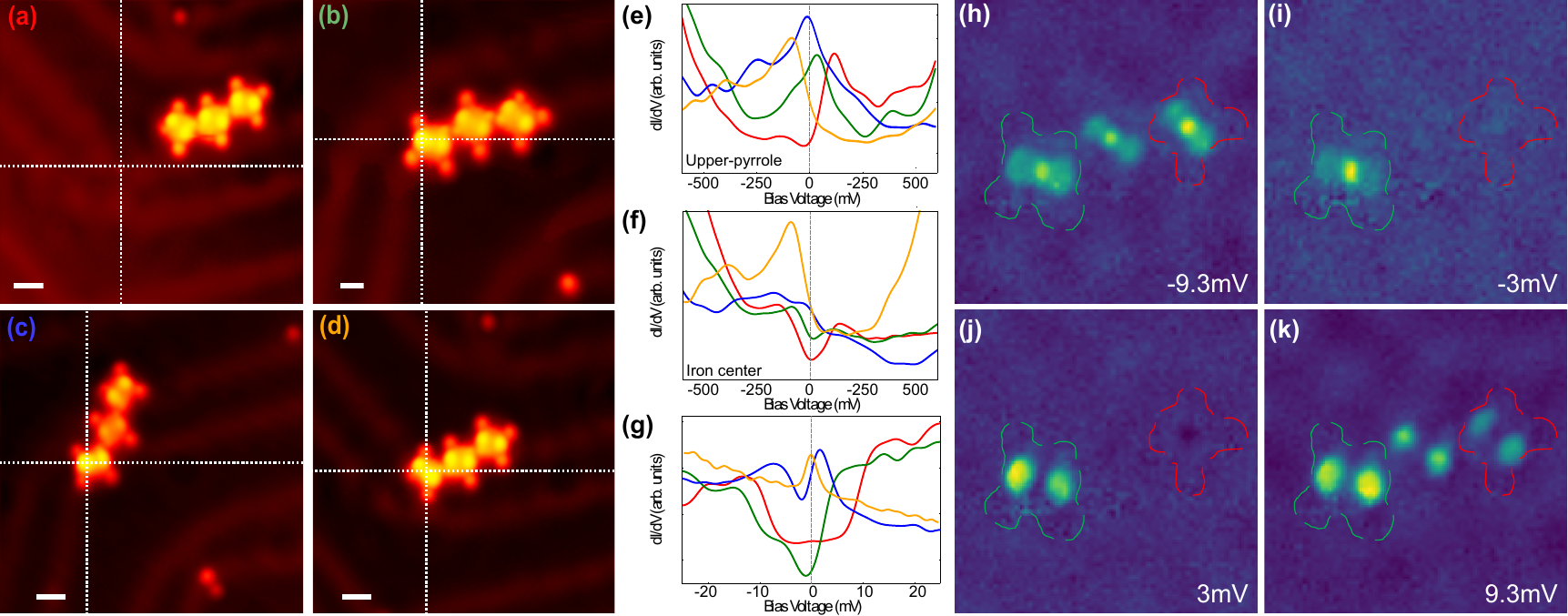}
  \caption{\textbf{Crossover from spin 1 nanomagnet to Kondo screen ground state through the mixed-valence regime}. \textbf{(a-d)} topography images showing the position of the first molecule of a chain with respect to the Br-decorated elbow site of the gold reconstruction. \textbf{(a)} the molecule is far from the Br site. \textbf{(b)} the upper pyrrole is centered on the Br-site. \textbf{(c)} the Br-site is located in between the center of the upper pyrrole and the Fe ion position. \textbf{(d)} the Fe atom is located just above the Br-site. \textbf{(e)} conductance spectra taken above the center of the upper pyrrole. \textbf{(f and g)} conductance spectra taken above the iron atom. red, green, blue and yellow colors correspond to the configurations of (a), (b), (c) and (d) respectively. \textbf{(h-k)} Background subtracted differential conductance images corresponding to various voltage recorded in the configuration of the case of image (b) (see Methods section). The left molecule is in the mixed-valence intermediate state (green dashed line) and the two other molecules behave as spin 1 nanomagnets (e.g. red dashed line). (h-k) voltages are $V = -9.3$\,mV, $V = -3.0$\,mV, $V = 3$\,mV and $V = 9.3$\,mV respectively. Experimental parameters: (a) \& (d) $V = 125$\,mV, $I=20 $\,pA, scalebar = $2$\,nm; (b) \& (c) $V = 200$\,mV, $I=100$\,pA, scalebar = $2$\,nm; (e) \& (f) $V_{stab}=800$\,mV, $I_{stab}=500$\,pA, lock-in parameters: $V_{m}=5$\,mV, $f=900$\,Hz; (g) $V_{stab}=30$\,mV, $I_{stab}=200$\,pA, lock-in parameters: $V_{m}=0.2$\,mV, $f=750$\,Hz; (h-k) $V_{stab}=30$\,mV, $I_{stab}=200$\,pA, size $12 \times 12$\,nm$^2$, lock-in parameters: $V_{m}=0.2$\,mV, $f=750$\,Hz. All spectra are normalized.}
  \label{fig:crossover}
\end{figure*}

The cases presented above are perfectly consistent with the scenario in which the molecule evolves from a quasi-atomic moment (spin 1) under the influence of magnetic anisotropy to an Fe atom carrying a screened magnetic moment resulting from a strong hybridization with the substrate.
Finally, the DFT calculations clearly indicate that the strength of the charge fluctuations should vary with the distance to the Br atom. Indeed the transition from spin 1 nanomagnet to Kondo screening, through the mixed-valence regime has been experimentally probed in a reversible manner and summarized in figure \,\ref{fig:crossover}.

\subsection{Crossover from spin 1 nanomagnet to Kondo-screened ground state through the mixed-valence regime}
It was possible to finely tune the electronic configuration of the iron atom by using a chain of molecules in order to stabilize positions for which a molecule deviates slightly from the Br site. The studied position of the molecules are presented in topography images (a-d) of figure\,\ref{fig:crossover} and the density of states close to the Fermi energy are measured by STS spectra figure\,\ref{fig:crossover} (g)). The spin 1 signature is observed  when the molecule is far from the elbow site as already mentioned above, figure\,\ref{fig:crossover} (a). When the molecule is positioned in such a way that the Br site is located below the upper pyrrole ring, image\,\ref{fig:crossover} (b), the LDOS shows a broadened and asymmetric step like shape originating from spin flip inelastic excitations but with a lower characteristic energy, indicating a renormalization of the anisotropy energy. At intermediate positions where the Br adsorption-site is located in between the Fe site and the upper pyrrole location, \,\ref{fig:crossover} (c), the spectroscopy shows a small anti-resonance at the Fermi level with a Fano-like shape which we believe is the extrapolation of the inelastic spin flip signature but for stronger charge fluctuations. When the molecule is strictly coinciding with the Br-adsorption site, figure\,\ref{fig:crossover} (d), it is exhibiting a Kondo peak.
Therefore, the situation of images\,\ref{fig:crossover} (b) \& (c) corresponds to two intermediate mixed-valence regimes between spin 1 nanomagnet and Kondo-screened ground states. Indeed as shown by D. Jacob in Ref.\,\cite{JacobD2018} the gradual increase of charge fluctuations alters the spectroscopic signatures of the spin 1 nanomagnet by renormalizing the magnetic excitations to lower energy. This gradually closes the gap and finally restores a Kondo peak as in the case of image\,\ref{fig:crossover} (d). In this process, the absence of particle-hole symmetry causes a typical Fano resonance to appear. The renormalization of the magnetic excitations is evidenced in pannels\,\ref{fig:crossover}\,(h-k) where differential conductance is used to map the inelastic spin-flip probability density in the real space in the same configuration as in (b). This situation corresponds to the occurrence of one molecule in the intermediate state and 2 molecules in the regime of the spin 1 nanomagnet. At $\pm 9.3$\,mV, spin-flip channels are open in all 3 molecules and the probability to induce a spin flip follows a two-lobes spatial pattern in all three molecules. At $\pm 3$\,mV, panels (i) \& (j) the spin-flip excitations are hindered by the magnetic anisotropy in the molecules far from the Br-site. On the contrary, the molecule in the intermediate regime shows spin-flip processes already being developed at $\pm 3$\,mV, indicating the renormalization of the magnetic excitation energy to a smaller scale.

This transition through a mixed-valence regime might be due to a net increase of Fe charge while remaining still in a regime of weak coupling with the substrate (our previously outlined scenario 2). Our DFT calculations rather suggest that the predominant effect is the drastic increase of the hopping from/to the substrate (scenario 1), instead. Accordingly, the spin crossover was found to be accompanied with an energy shift of the frontier orbitals which are localized at the upper pyrrole rings (frontier orbital A, FOA) and at the iron atom position (frontier orbital B, FOB). A carefully analysis of figures\,\ref{fig:crossover}\,(e) \& (f) allows to follow the evolution of FOA and FOB energies when moving the molecule. FOA shows a continuous evolution of density of states from the unoccupied states towards occupied states when approaching the molecule to the Br-site, indicating an increase of the orbital occupation. FOB shows a reverse behavior with a shift from below the Fermi level towards the unoccupied states.
The gap between the frontier orbitals is reducing when approaching the Br site so the d level splitting tends to be weakened and could be favoring a decrease of orbital momentum and a renormalization of the magnetic anisotropy in agreement with the closing of the steps of conductance due to spin-flip excitations in the measured data of panel\,\ref{fig:crossover} (g) and figures\,\ref{fig:crossover} (h-k).

Following the work in ref.\cite{RubioVerduC2018}, the conductance images presented figure\,\ref{fig:Chains} (c-e) support the hybrid Fe 3d$_{\rm xz}$, d$_{\rm yz}$ character of FOA and the strong Fe 3d$_{z^2}$ character of FOB in agreement with the orbital occupations in the DFT calculations (figure\,\ref{fig:DFT1} (c)). It is interesting to note that according to conductance images of figure\,\ref{fig:crossover} (h-k), the spin-flip probability density is following the 3d$_{\rm xz}$, d$_{\rm yz}$ symmetry at energies above and below the Fermi energy (two-lobes shape). On the contrary, the 3d$_{z^2}$ character of the excitations (bright localized spot at the Fe site) is only visible at negative bias, so for occupied states. This difference is explaining the breaking of the electron-hole symmetry observed in inelastic spectra.

The fano-like shape of the blue spectrum in figure\,\ref{fig:crossover} (g) occurs when the frontier orbitals are overlapping at the Fermi level, as visible in figure\,\ref{fig:crossover} (e) \& (f) and when the charge fluctuations between orbitals are expected to be strong. The hypothesis of intermediate regimes driven by charge fluctuations is well described by prediction in ref.\,\cite{JacobD2018} as well as the shape of the spectrum. However in the present case the frontier orbitals, could also recover their degeneracy when overlapping in the intermediate regimes, implying a possible Kondo SU(4) effect\,\cite{Maki2012,MinamitaniE2012}, although this is less likely in a molecular system\,\cite{KuegelJ2018a} and cannot be easily distinguished from the simple effect of strong charge fluctuations\,\cite{Fern_ndez_2015}.
% These charge fluctuation driven intermediate regimes follows well the prediction of ref.\,\cite{JacobD2018} where a spin 1 impurity under charge fluctuation is studied, yet the recovering of degeneracy of frontier orbitals in intermediate regimes could be possible, implying a SU(4) Kondo effect \cite{Maki2012,MinamitaniE2012} though it is less likely to occur in molecular system \cite{KuegelJ2018a} and cannot easily be distinguished from the simple effect of strong charge fluctuations\,\cite{Fern_ndez_2015}. 
Finally, when the iron atom is located above the Br site, FOA migrates below the Fermi level and FOB is washed out towards incoherent excitation and only the Kondo resonance survives at $\rm{E_F}$. 

\section{Conclusion}
We have revealed in experiments and DFT calculations that the spin state switching is caused by the strong hybridization potential at Br-decorated elbow sites which induce charge fluctuation and a renormalization of the magnetic anisotropy when molecules are adsorbed on these sites. 

The direct consequence of charge fluctuations is to tune the energies of two molecular frontier orbitals which are respectively closest to the Fermi level in positive and negative energies. The energetic overlapping of these two frontier orbitals gives rise to low-energy spin excitation channels near Fermi level. Under the condition of relative small charge fluctuations, the two frontier orbitals are far from Fermi level and the overlapping is small, hence the molecule will have high spin (S = 1) state, while under strong charge fluctuation these two orbitals can cross Fermi level, resulting a renormalization of the magnetic anisotropy and ultimately the Kondo screened ground state.

Our work provides a new approach for tuning not only the spin state but also the intermediate valence character/charge fluctuation in hybrids made of molecules on reconstructed surfaces. 

\section{Methods}
Iron(III) 5,15-(di-4-bromophenyl)-10,20-(di-4-pyridyl)porphyrin 
chloride molecules molecules were ordered from PorphyChem company (95\% purity, CAS not assigned). 
Samples were prepared under ultrahigh vacuum (base pressure $1 \times 10^{-10}$ mbar) just before their scanning tunneling studies.
The gold substrate was prepared by repeated cycles of Ar ion sputtering followed by annealing at 900 K. 
The molecules were deposited by organic molecular beam epitaxy (OMBE) when the sample was kept at Room Temperature (T$_{crucible}$=575K). It was observed that the molecule source is also depositing Br atoms on the surface of gold. We have attributed this concomitant deposition of Br with $\rm{Fe-DPyDBrPP}$ to the stochastic Br detaching from $\rm{Fe-DPyDBrPP}$ molecules inside the crucible when heated at evaporation temperature. In-crucible Br detaching form similar molecules was already reported in ref.\,\cite{grill2007nano} at 590K. In our case, the low amount of deposited Br atoms was seen to be enough for occupying all the elbow sites of the herringbone reconstruction.
Once the molecules deposited, the sample temperature was quenched at liquid nitrogen temperature then at helium liquid temperature in less than 20 minutes. For low coverage this procedure produces single monomers scattered on the surface. 
A post annealing at 400 K for 10 minutes was used for producing molecular chains by Ullman's coupling at the surface following recipes in ref.\,\cite{grill2007nano,KrasnikovSA2011,DiLulloA2012,LafferentzL2012,ClairS2019}. As for other porphyrins deposited on surfaces, the chlorine atom is detached from the molecule during the deposition or the post-annealing. 

STM/STS measurements were achieved in-situ in ultrahigh vacuum by means of a modified "Tyto SPM" from SPECS\textsuperscript{TM} at a base temperature of 1.3\,K; the tunneling bias was applied to the sample. Mechanically cut PtIr tips were used. Soft lateral manipulation as follows was used for moving the molecules 1) positioning the tip above the molecule to move 2) Turning off the microscope feedback loop and approaching the tip towards molecule in order to increase tip-molecule interaction 2) Moving the tip to the destination in constant height mode 3) Retracting the tip and engaging the feedback loop.
The analyses of the microscopy and spectroscopy data were done with WSxM software\,\cite{WSXM2007} and homemade python procedures respectively\,\cite{GaoYZ2022}. 

The fit of spectra from figure\,\ref{fig:Singles} \& \ref{fig:Chains} involved a large Gaussian background mimicking the frontier orbital DOS plus a specific function associated to spin-flips (step functions introduced in \cite{lambe1968molecular}) or Kondo ground states (Frota function). Marginally, a convolution with a Gaussian function was used to clean up the incoherent noise of all conductance spectra presented here.

The spin-polarized Density-Functional Theory (DFT) calculations have been carried out with v.$\sim$6.8 of the \textsc{Quantum ESPRESSO} package\,\cite{Giannozzi_2009,Giannozzi_2017} within the Projector Augmented-Wave (PAW) scheme\,\cite{Blochl_1994} and using the Perdew-Burke-Ernzerhof (PBE) functional\,\cite{Perdew_1996}.
PAW pseudopotentials have been taken from the PSlibrary\,\cite{DalCorso_2014,PSlibrary,QEpseudo}. 

The Au(111) surface has been modeled by a finite slab consisting of 3 gold layers with $15$\,\AA of vacuum between periodic replicas along the $z$ direction. The Au-Au nearest-neighbor distance has been set to $2.93$\,\AA, the PBE equilibrium distance of the corresponding fcc bulk structure, which is $\lesssim 2\%$ larger than the experimental room-temperature value\,\cite{Davey_1925}. 

Given the sizeable extension of the molecule, a $10 \times 10$ supercell of Au atoms on the $xy$ plane has been used in order to avoid interaction between its periodic replicas, for a total of 300 Au atoms in the simulation cell.
The molecule geometry has been relaxed in the vacuum until all components of the forces were smaller than $10^{-3}$\,Ry/$a_{\rm B}$, with $a_{\rm B}$ being the Bohr radius.
Fe Orbital polarizations and Density of states projected over the atomic states of the Fe pseudopotential can be found in supporting information.
The projected density of states (pDoS) and the Mulliken orbital occupations and polarizations have been evaluated by projecting the Kohn-Sham orbitals on the atomic states of the pseudopotentials. They have been computed by considering different adsorption geometries between the molecule and three different substrates: the clean Au(111) surface, the Au(111) surface with a Au impurity and the Au(111) surface with a Br impurity.
The impurities have been placed at the preferred PBE adsorption site at the equilibrium distance, which has resulted in the hollow-hcp site for both the Au and the Br impurity, respectively at $2.16$\,\AA and $2.23$\,\AA from the Au(111) surface.
All calculations have been performed with and a plane-wave cutoff of $\epsilon_{\rm cut} = 40$\,Ry on the wavefunctions and 320\,Ry on the density. Brillouin zone integrals have been evaluated at the $\Gamma$ point using a gaussian smearing of $\sigma = 0.01$\,Ry.
The hopping parameters have been computed via the projection method as implemented in the {\sc Wannier90} package\,\cite{Pizzi_2020}.

\section{Acknowledgement}
 Cheryl Feuillet-Palma is acknowledged for her constructive discussions. TG thanks Paolo Restuccia for fruitful discussions. LdM acknowledges Giorgio Sangiovanni for useful discussions. SV, DR and SP acknowledge the French National Research Agency for the support of the SHOGUN project, Re. ANR-22-CE09-0028-01. TG and LdM are supported by the European Commission through the ERC-CoG2016, StrongCoPhy4Energy, GA No724177.

%\bibliography{References_05_10}

% \newpage
% \begin{figure}
%   \centering
%   \includegraphics[width = 9 cm]{TOC.pdf}
%   \caption*{Table of Contents Graphics}
%   \label{TOC}
% \end{figure}
\end{document}